\documentclass[aps,prd,natbib,twocolumn,groupedaddress,nofootinbib,preprintnumbers,superscriptaddress,nolongbibliography,secnumarabic]{revtex4-2}  
\usepackage{graphicx}
\usepackage{amsmath}
\usepackage{amsfonts}
\usepackage{bm,bbm}
\usepackage[colorlinks=true,citecolor=red,linkcolor=blue,breaklinks=true]{hyperref}
\usepackage[figure, table]{hypcap}
\usepackage{amssymb}
\usepackage{xcolor}
\usepackage{subfigure}
\usepackage{setspace}
\usepackage{footnote}
\usepackage{multirow}
\usepackage[normalem]{ulem}
\usepackage[utf8]{inputenc}
\usepackage{mathrsfs}
\usepackage{slashed}
\usepackage{longtable}
\usepackage{enumitem}
\usepackage{cancel}
\usepackage{booktabs}
\usepackage{blindtext}

\usepackage{hhline}

\newcommand{\neff}{\ensuremath{N_{\text{eff}}}\xspace}

\newcommand{\s}{\ensuremath{\text{ s}}\xspace}

\usepackage{orcidlink}
\newcommand{\orcid}[1]{\,\orcidlink{#1}}

\usepackage{xcolor}
\definecolor{linkcolor}{rgb}{0.0, 0.47, 0.75}
\definecolor{citecolor}{rgb}{1.0, 0.5, 0.0}
\hypersetup{
  linkcolor  = linkcolor,
  citecolor  = linkcolor,
  urlcolor   = linkcolor,
  colorlinks = true
}

\begin{document}
\preprint{CERN-TH-2026-056}
\title{What does it take to have $N_{\rm eff} < 3$ at CMB times?}

\author{Miguel Escudero Abenza\orcid{0000-0002-4487-8742}}
\email{miguel.escudero@cern.ch}
\affiliation{Theoretical Physics Department, CERN, 1 Esplanade des Particules, CH-1211 Geneva 23, Switzerland}

\author{Maksym Ovchynnikov\orcid{0000-0001-7002-5201}}
\email{maksym.ovchynnikov@cern.ch}
\affiliation{Theoretical Physics Department, CERN, 1 Esplanade des Particules, CH-1211 Geneva 23, Switzerland}

\author{Neal Weiner\orcid{0000-0003-2122-6511}}
\email{neal.weiner@nyu.edu}
\affiliation{Center for Cosmology and Particle Physics, Department of Physics, New York University, New York, NY 10003, USA}

\begin{abstract}
\noindent

The vast majority of extensions of the Standard Model affecting the number of effective relativistic neutrino species ($N_{\rm eff}$) do so additively, namely, they enhance this quantity with some light state contributing to dark radiation. In this work, we consider precisely the opposite case: new physics scenarios that can lead to $N_{\rm eff} < 3$ that are consistent with all known cosmological, astrophysical, and laboratory data. We are motivated by three main reasons: 1) a recent measurement from  ACT and SPT in combination with Planck that leads to $N_{\rm eff} = 2.81\pm0.12$, 2) by a new and powerful measurement of the primordial helium abundance, which anchors $N_{\rm eff}$ to be very close to the Standard Model value one second after the Big Bang, 3) by the deployment of the Simons Observatory which will provide precise tests of the radiation content in the Universe and which may detect with a high significance cosmologies with $N_{\rm eff}<3$. We survey the main theoretical possibilities and find that only a few simple scenarios can consistently give $N_{\rm eff}=2.81\pm0.12$. One class consists of thermal electrophilic relics with masses $m\sim 8$--$13\,{\rm MeV}$. Another consists of out-of-equilibrium particles decaying to $e^+e^-$ or $\gamma\gamma$, with a rather particular lifetime $0.05\,{\rm s}\lesssim \tau \lesssim 3\,{\rm min}$, mass $250\,{\rm MeV}\lesssim m \lesssim 600\,{\rm MeV}$, and abundance $\rho/\rho_\gamma\sim 0.1$ at decay. Thermal electrophilic particles are especially interesting because they can account for the dark matter in the Universe and can be tested in experiments such as SENSEI, DAMIC-M, and Oscura, and their portals to the visible sector at experiments such as NA64 and LDMX. We conclude that if the Simons Observatory confirms that $N_{\rm eff} \simeq 2.8$, it will point to very specific extensions of the Standard Model. 
\end{abstract}

\maketitle

\section{Introduction}

One second after the Big Bang, weak interactions freeze out, and the Universe contains at least three types of relativistic species: electrons/positrons, photons, and neutrinos. As the Universe cools and the temperature drops below the electron mass, $e^+e^-$ pairs annihilate mostly into photons. This heats the photons relative to the neutrinos and leads to the familiar ratio $T_\gamma/T_\nu \simeq 1.4$ between the temperatures of the Cosmic Microwave Background (CMB) and the Cosmic Neutrino Background. CMB observations are sensitive to this ratio via the number of effective relativistic neutrino species:
\begin{subequations}\label{eq:Neffdef}
\begin{align}
N_{\rm eff} &\equiv \frac{8}{7} \left(\frac{11}{4}\right)^{4/3} \left(\frac{\rho_{\rm rad}-\rho_\gamma}{\rho_\gamma}\right)\,,\\
		&\to \frac{8}{7} \left(\frac{11}{4}\right)^{4/3} \left(\frac{\rho_{\rm \nu}}{\rho_\gamma}\right)\,\,\,\, \text{(if only } \nu \text{ \& } \gamma) \,,
\end{align} 
\end{subequations}
where this quantity should be evaluated at $T \lesssim 10\,{\rm eV}$ somewhat before recombination. In the second line, we have particularized the expression for the case where only neutrinos and photons are present as relativistic species. The value of \neff has been computed very precisely in the literature~\cite{Akita:2020szl,Froustey:2020mcq,Bennett:2020zkv,Jackson:2023zkl,Jackson:2024gtr,Ihnatenko:2025kew,Escudero:2025kej} and the consensus Standard Model prediction is $N_{\rm eff}|_{\rm SM}=3.044$, which we will often round to 3 for simplicity. This simple quantity is a central test of the Standard Model in cosmology, with important implications both for the primordial helium and deuterium abundances (probing \neff in the 100 keV -- 1 MeV era) and the CMB (probing \neff in the 1 -- 10 eV era). 

There is an array of extensions of the Standard Model which predict contributions to $N_{\rm eff}$ in some form of dark radiation, making $N_{\rm eff}$ measurements a key probe of physics beyond the Standard Model. Canonical candidates for this dark radiation include eV-scale sterile neutrinos~\cite{Barbieri:1990vx,Gariazzo:2019gyi}, hot axions~\cite{Chang:1993gm,Hannestad:2005df,Caputo:2024oqc}, majorons~\cite{Chikashige:1980ui,Sandner:2023ptm}, gravitinos~\cite{Pagels:1981ke}, Goldstone bosons~\cite{Weinberg:2013kea,Lindner:2011it}, and many more, see e.g. for a review~\cite{Allahverdi:2020bys}. Improving early Universe constraints on \neff can thus provide key insights into a wide range of beyond the Standard Model (BSM) scenarios.

In this context, in January 2026, a new and very precise measurement of the primordial helium abundance was reported as a key output of the LBT $Y_{\rm P}$ project~\cite{Aver:2026dxv}:
\begin{align}\label{eq:Ypmeasurement}
Y_{\rm P} = 0.2458 \pm 0.0013\,.
\end{align}
High-precision measurements of the primordial helium abundance are obtained from ionized extragalactic metal-poor galaxies. The collaboration in Refs.~\cite{Skillman:2026ltj,Aver:2026dxv} obtained new high-quality spectra for about 50 such systems, improved the modeling of their emissivities, and performed a global analysis. The result is a 0.5\% determination of the primordial helium abundance, in agreement with the Standard Model, and is more precise than any previous measurement. 

The primordial helium abundance produced during Big Bang Nucleosynthesis (BBN) is highly sensitive to the expansion rate of the Universe around one second after the Big Bang. Modeling the expansion rate of the Universe by a simple re-scaling of the neutrino energy density $\rho_{\nu}(t)\to \rho_{\nu}^{\rm SM}(t) \cdot N_{\nu}/3$, it has been shown that this implies: $N_{\nu} = 2.93 \pm 0.08 $~\cite{Yeh:2026pil}. This provides an anchor for $N_{\rm eff}$ in the very early Universe with remarkable precision, at a value near that of the Standard Model.

The CMB provides an additional key anchor at later times. In particular, through its effect on the damping tail of the CMB, the presence of additional radiation in the Universe can be tested. Recently, the Atacama Cosmology Telescope (ACT) and the South Pole Telescope (SPT), in combination with Planck data, reported in 2025 the most precise measurement of $N_{\rm eff}$ to date~\cite{AtacamaCosmologyTelescope:2025nti,ACT:2025fju,SPT-3G:2025bzu}:
\begin{align}\label{eq:NeffCMB2025_}
    N_{\rm eff} = 2.81 \pm 0.12\,\, \text{[ACT+SPT+Planck]}\,.
\end{align}
This result, which is approximately two sigma below the SM value, already strongly constrains new sources of relativistic particles in the Universe\footnote{In this context, we shall also mention that an independent analysis~\cite{Tristram:2025you} reports a somewhat larger inferred value for $N_{\rm eff}$ using a very similar data set combination. The origin of this higher value compared to that reported by the collaborations is unclear.}.

The Simons Observatory (SO)~\cite{SimonsObservatory:2018koc} is already operational and aims to reach a one sigma precision of $\sigma_{\neff}\simeq 0.045$ \cite{SimonsObservatory:2025wwn}. In light of existing constraints, it is clear that a five sigma -- or even three sigma -- discovery of additional relativistic radiation by SO is already strongly disfavored by existing data. This situation is driven by the existing preference for $\neff<3$.

On the other hand, a significant negative deviation of \neff is still consistent with the data and could be conclusively tested by SO. Given this, it is clearly important to understand scenarios which yield $N_{\rm eff}<3$. In this work, we shall focus our attention on scenarios where this is achieved, with the hope of understanding a range of models that might be discovered by the Simons Observatory. 

Such models are not new considerations, and many scenarios are known to lead to $N_{\rm eff} < 3$, and examples include thermal electrophilic species~\cite{Kolb:1986nf,Serpico:2004nm,Ho:2012ug,Boehm:2013jpa,Nollett:2013pwa,Sabti:2019mhn,Depta:2019lbe,Giovanetti:2021izc,Chu:2022xuh,Depta:2020wmr,Escudero:2025avx}, relics decaying into various Standard Model species~\cite{Masso:1995tw,Sabti:2020yrt,Boyarsky:2021yoh,Rasmussen:2021kbf,Ovchynnikov:2024xyd,Ovchynnikov:2024rfu,Akita:2024nam,Cadamuro:2011fd,Millea:2015qra,Depta:2020wmr,Escudero:2025avx}, and low-reheating temperature scenarios~\cite{Kawasaki:1999na,Kawasaki:2000en,Giudice:2000ex,Hannestad:2004px,deSalas:2015glj,Hasegawa:2019jsa,Barbieri:2025moq}. Our goal is to explore these scenarios in the precision cosmology context in which we live, especially in concert with the new primordial helium measurement in Eq.~\eqref{eq:Ypmeasurement}. 

\begin{figure*}[!t]
\centering
\hspace{-0.cm}\includegraphics[width=0.96\textwidth]{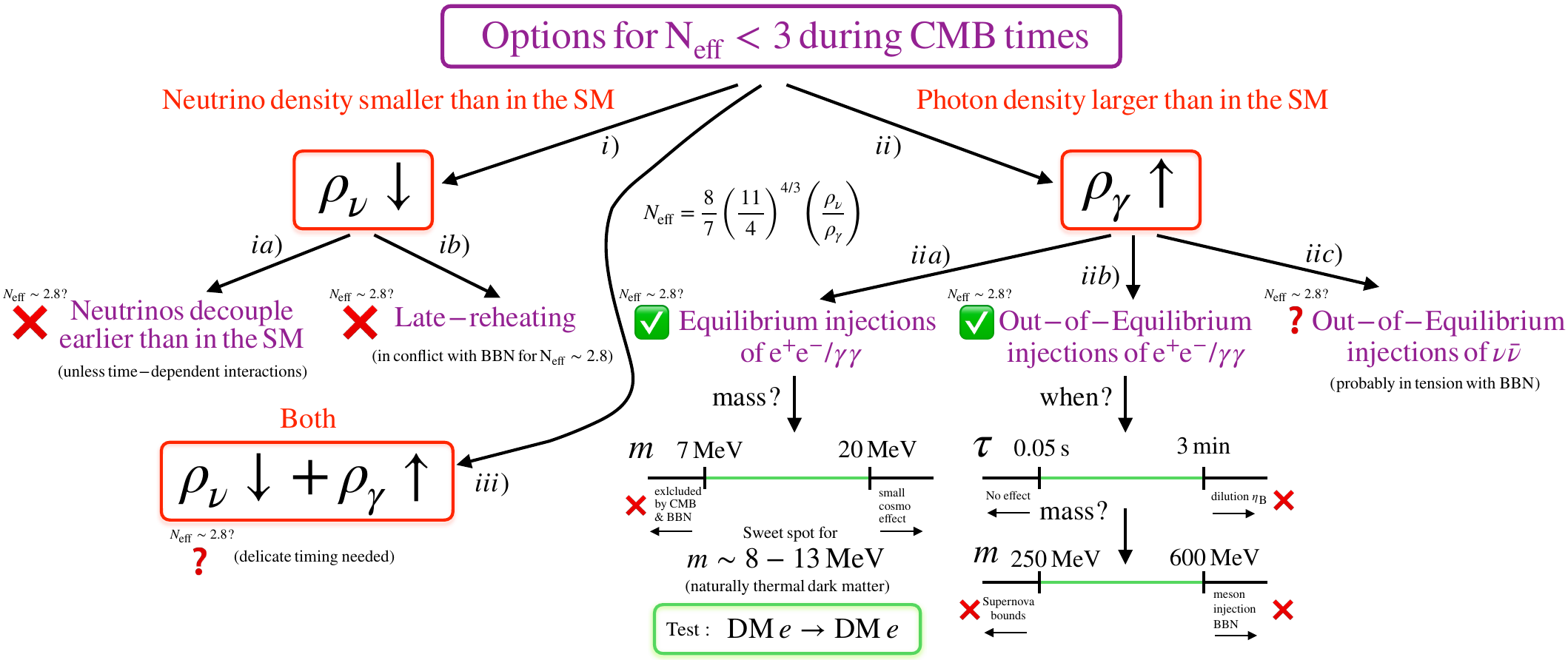}
\caption{
Summary of the various options from Section~\ref{sec:globalconsiderations} that can lead to $N_{\rm eff} < 3$ at CMB times. We highlight those which may lead to $N_{\rm eff} \simeq 2.8$ and hence to a good fit to current CMB data, as well as to allow a potential $\sim 5\sigma$ detection by the Simons Observatory (SO). As discussed in the text, two clear and simple options are iia) thermal electrophilic species with masses around $m\sim (8-13)\,{\rm MeV}$ (see Fig.~\ref{fig:thermal}) and which may be tested with dark matter-electron scatterings on Earth, and iib) out-of-equilibrium injections of $e^+e^-/\gamma \gamma$ from particles with lifetimes $\tau_X \sim (0.05-100)\,{\rm s}$ and masses $m_X\sim (250-600)\,{\rm MeV}$, see Section~\ref{sec:nonthermalcase}.
}
\label{fig:summary}
\end{figure*}

In particular, we will discuss various physical possibilities that can lead to $N_{\rm eff}<3$. We will focus on those models where $N_{\rm eff}$ may be as low as 2.8 and still be compatible with the rest of known cosmological, astrophysical, and laboratory data. In particular, a model capable of lowering $N_{\rm eff}$ down to the currently favored 2.8 central value can be discovered with $\sim 5\sigma$ significance by the Simons Observatory. As a consequence, we will take this value as a key interesting benchmark for discovery. 

The remainder of this work is structured as follows: In Section~\ref{sec:globalconsiderations}, we present an overview of the main phenomenological avenues to have $N_{\rm eff} < 3$. Then in Sections~\ref{sec:thermaleq} and~\ref{sec:nonthermalcase} we look in detail at two particularly simple set-ups that we identify can do the job: thermal electrophilic species and decaying axion-like particles or dark photons, respectively. We study their parameter space and highlight the region where $N_{\rm eff} \simeq 2.8$ while being consistent with all other known data. We draw our conclusions in  Section~\ref{sec:conclusions}, highlighting how some of these scenarios may be tested in the future.

\section{General considerations: What are the options to make $N_{\rm eff}<3$?}\label{sec:globalconsiderations}

$N_{\rm eff}$ as relevant for CMB observations is defined explicitly in Eq.~\eqref{eq:Neffdef}. To reduce $N_{\rm eff}$ below its Standard Model value, there are three logical possibilities: i), one can lower $\rho_\nu$ relative to $\rho_\nu|_{\rm SM}$, which we denote by $\rho_\nu\downarrow$; ii), one can raise $\rho_\gamma$ relative to $\rho_\gamma|_{\rm SM}$, which we denote by $\rho_\gamma\uparrow$; iii), one can do both simultaneously, which we denote by $\rho_\nu\downarrow+\rho_\gamma\uparrow$. We begin with general considerations on whether these are viable possibilities, referring to Fig.~\ref{fig:summary} for a quick visual summary.

\subsection{Reducing the energy density in neutrinos}

To assess whether $\rho_\nu\downarrow$ is viable, we first review neutrino decoupling. In the usual radiation-dominated early Universe picture, thermal equilibrium is established at very early times and as the Universe cools down, eventually, at $T\sim 2\,{\rm MeV}$ the electroweak interactions of neutrinos freeze out and from then on they simply free-stream. Therefore, reducing the neutrino energy density relative to photons requires modifying the decoupling process. The possibilities are: ia) making neutrinos decouple earlier (at higher temperatures), e.g., if neutrinos were to decouple before muons annihilate, then $N_{\rm eff} < 1.6$, as a significant fraction of the entropy density of the Standard Model particles would end up in photons and not in neutrinos, or ib) by not producing a thermal plasma of neutrinos.

Option ia) points towards non-standard neutrino interactions. It is in tension with experimental observations unless the coupling strengths controlling the interactions between neutrinos and charged leptons are time-dependent. Option ib) is viable in principle. It has been explored in cosmologies with a very low reheating temperature~\cite{Kawasaki:1999na,Kawasaki:2000en,Giudice:2000ex,Hannestad:2004px,deSalas:2015glj,Hasegawa:2019jsa,Barbieri:2025moq}, close to $T\sim {\rm MeV}$, when neutrinos decouple. In these models, one assumes that an out-of-equilibrium particle decays and generates the thermal Universe. For example, if the reheating field decays electromagnetically, $\Phi\to e^+e^-$, it initially populates only the electromagnetic plasma. Neutrinos are then produced secondarily through processes such as $e^+e^-\to \nu\bar{\nu}$. If the electromagnetic plasma reaches only $T_{\rm reh}\sim {\rm MeV}$, neutrinos do not fully thermalize. Their energy density is then smaller than in the Standard Model, which leads to $N_{\rm eff}<3$. 

However, the issue with these models is Big Bang Nucleosynthesis. In a Universe with $N_{\rm eff} \simeq 2.8$ as obtained in these models, the primordial element abundances of both deuterium and helium would be in strong tension with observations (see, e.g., Fig. S1 in~\cite{Barbieri:2025moq}). 
In this context, one can also consider the possibility of bleeding energy from the neutrino bath {\em after} BBN. However, if the energy goes into a new, invisible radiation, this will not have any consequence for CMB measurements of \neff, which measures the totality of invisible, non-interacting radiation. Thus, for such a scenario to be effective, the only option would be to deplete $\rho_\nu$ and to simultaneously increase $\rho_\gamma$, which we shall discuss shortly.

\subsection{Enhancing the energy density in photons}

Option ii), having more energy density in photons, is, on the other hand, possible, and we will focus on it. Three possibilities can lead to this: iia) a thermal injection of photons or $e^+e^-$ (as they thermalize), iib) a similar type of injection but from a particle that is decaying out-of-thermal equilibrium, and iic) an injection of out-of-equilibrium neutrinos which then annihilate into $e^+e^-$ neatly producing more electromagnetic than neutrino radiation. 

The viability of these scenarios depends crucially on when the energy is injected. Two considerations determine the relevant time window. First, neutrinos remain in thermal equilibrium down to $T\sim 2\,{\rm MeV}$, corresponding to $t_U\sim 0.1\,{\rm s}$. Therefore, any electromagnetic injection at much earlier times is efficiently shared with the neutrino sector, for example, through $\Phi\to e^+e^-$ followed by $e^+e^-\to \nu\bar{\nu}$. In addition, the primordial abundance of deuterium is very sensitive to the baryon density of the Universe at the time of nucleosynthesis, ${\rm D/H}|_{\rm P} \propto (\Omega_b h^2)^{-1.6}$. The prediction of the deuterium abundance still suffers from somewhat large uncertainties from nuclear reaction rates, but one can infer from it a baryon density with an uncertainty of at least $\sim 3\%$~\cite{Pisanti:2020efz,Gariazzo:2021iiu}. Critically, this inference is in overall good agreement with the value of the baryon density obtained from CMB observations, and this means that it cannot change substantially between BBN and recombination, see e.g.~\cite{Yeh:2022heq}. Since BBN took place at $t_U \sim 3\,{\rm mins}$ and an injection of energy leading to $\Delta N_{\rm eff} = -0.2\pm 0.1$ would modify the baryon density by $\simeq (7\pm 3.5)\%$, it is clear that the injection needs to take place within a very specific window:
\begin{align}
\label{eq:t_window}
0.05\,{\rm s} \lesssim t_U\lesssim 100\,{\rm s}\,.
\end{align}
Of course, injections of electromagnetic radiation at $t_U \gtrsim 100\,{\rm s}$ may also be further constrained by nuclear photodisassociation for $t_U \gtrsim 5\times 10^{4}\,{\rm s}$~\cite{Kawasaki:1994sc} and CMB spectral distortions for $t_U \gtrsim 10^6\,{\rm s}$~\cite{Hu:1992dc,Poulin:2016anj}, see also~\cite{Sobotka:2022vrr}.

Having established this timeline, there are three distinct types of options for the injection: iia) it may be thermal (i.e., a particle annihilating or decaying into $e^+e^-/\gamma\gamma$ in thermal equilibrium with the plasma), iib) it may be non-thermal (where the decay happens out-of-equilibrium), and iic) it may actually be into neutrinos. Some considerations are in order. For the thermal injections iia), it is clear that the phenomenology will be dictated primarily by the given particle mass. Since neutrino decoupling happens at $T\sim 2\,{\rm MeV}$, the particle would need to have a mass somewhat close to but larger than this scale. If it is much smaller, then it would lead to heavily distorted values for the primordial element abundances and $N_{\rm eff}$, see e.g.~\cite{Sabti:2019mhn}. We quantitatively investigate this scenario explicitly in Section~\ref{sec:thermaleq}. The non-thermal case iib) features similar phenomenology, and we will investigate it explicitly in Section~\ref{sec:nonthermalcase}. 

Finally, let us comment on out-of-equilibrium injections of $\nu\bar{\nu}$ with $E_\nu\gg 3.15\,T$. Somewhat counterintuitively, such injections can also lead to $N_{\rm eff}<3$. It turns out that neutrinos, scattering against a thermal bath of $e^+e^-$ and $\nu\bar{\nu}$ with a similar temperature, may lead to more energy going to $e^+e^-$ effectively due to the enhanced cross section for $\nu\bar{\nu}\to e^+e^-$ processes as compared to the joint effect of $\nu\bar{\nu}\to \nu\bar{\nu}$ and $e^+e^-\to \nu\bar{\nu}$. This feature has been observed in several studies that considered scenarios with neutrinophilic particles~\cite{Hannestad:2004px,Boyarsky:2021yoh,Rasmussen:2021kbf,Ovchynnikov:2024xyd,Ovchynnikov:2024rfu} or species decaying into muons and pions~\cite{Akita:2024nam}, for the (particle-mass-dependent) lifetime range $\tau \lesssim 1\text{ s}$.

Although this mechanism can, in principle, lower \neff, the same high-energy neutrinos also interact efficiently with protons in the plasma. This generically enhances the helium abundance through processes such as $\bar{\nu}_e+p\to n+e^+$. While no dedicated studies yet exist~\cite{Akita:2026xwu}, for $N_{\rm eff}\sim 2.8$ it is likely that BBN excludes this possibility, and we will not consider it further in our study.\footnote{More complicated cosmological histories combining neutrinophilic decays with other phenomena may lead to a negative $\Delta \neff$ without spoiling BBN~\cite{Akita:2026gee}.}

\subsection{Reducing the energy density in neutrinos and enhancing the one in photons}

We now turn to option iii), in which the neutrino energy density is reduced while the electromagnetic energy density is increased. One can imagine a scenario in which, soon after neutrino decoupling at $T\sim 2\,{\rm MeV}$, neutrinos produce a new state $\chi$ through oscillations or other interactions. This state then decays out of equilibrium into neutrinos and electromagnetic radiation. The decay must occur out of equilibrium; otherwise, the neutrino and photon temperatures would re-equilibrate, typically yielding $N_{\rm eff}>3$. Considering this decay to be out-of-equilibrium would lead to $N_{\rm eff}<3$ since this would allow a net energy transfer from the neutrinos to photons. There are mechanisms where neutrinos thermalize via oscillations with a self-interacting dark sector~\cite{Aloni:2023tff,Giovanetti:2024orj} or just by annihilations~\cite{Berlin:2017ftj}. The question becomes how to transfer energy back to the photons. One may envision doing it by an out-of-equilibrium decay of the type $\chi \to \nu + \gamma$, as featured for instance by right-handed neutrinos with magnetic moments~\cite{Brdar:2020quo}. 

In the simplest models of this type, a natural tension arises: a large primordial population of $\chi$ states would be produced early by the same interactions that lead to its decay. With a long lifetime, when their decay produces roughly equal electromagnetic and neutrino energy densities, and because in the Standard Model $\rho_\nu < \rho_{\gamma}+\rho_e$ at all epochs, the net decay would lead to $N_{\rm eff} >3$, see~\cite{Brdar:2020quo}. However, relatively simple modifications can evade this. In the presence of a dark sector self-interaction, the effective mixing angle can be temperature-dependent~\cite{Dasgupta:2013zpn,Chu:2015ipa}. This may suppress the early production of sterile states, such that they are only produced after neutrino decoupling~\cite{Aloni:2023tff}. As such, the energy in the sterile states comes from the neutrino sector only. When they subsequently decay into both sectors, this now pushes \neff down. However, such a model requires a rather elaborate conspiracy of masses and timescales. We shall not pursue it here, but note it is a viable possibility worthy of study should $N_{\rm eff}< 3$ be robustly found.

\subsection{Summary}

We presented in this section what we believe are the main avenues to have $N_{\rm eff}<3$ at CMB times, and which we summarize in Fig.~\ref{fig:summary}. Out of the various possibilities, two viable ones appear as particularly simple and have the potential to lead to $N_{\rm eff}$ as small as 2.8. In what follows, we study their CMB and BBN phenomenology explicitly. These are particles that inject electromagnetic radiation in the window dictated by Eq.~\eqref{eq:t_window}, i.e., cases iia) and iib).  In Table~\ref{tab:models} we summarize the two scenarios with the key benchmark models that reproduce them.

\begin{table*}[t]
\renewcommand{\arraystretch}{1.2}
\setlength{\arrayrulewidth}{.25mm}
\centering
\small
\setlength{\tabcolsep}{0.18 em}
\begin{tabular}{ c | c | c  }
\hline\hline
\textbf{Scenario}$\,$       	&	\textbf{Specific Model Realization}$\,$ 	&	\textbf{Interaction Lagrangian}$\,$         \\ \hline
$\,\,$ \multirow{2}{*}{{iia) Thermal electrophilic particles}$\,$}    $\,\,$   	&$\,\,$	Complex scalar dark matter, $\phi$  $\,\,$	&	\multirow{2}{*}{$\,$ $\mathcal{L} = -i g_DA'_\mu (\phi^\star \partial^\mu \phi - \phi\partial^\mu \phi^\star)-\epsilon e A_\mu' J_{\rm em}^\mu$ $\,$ }         \\ 
&$\,\,$	interacting with a kinetically mixed $A'$  $\,\,$	&	      \\ \hline
iib) Out-of-equilibrium decaying         	&	Photophilic axion-like particle, $a$ 	&	$\mathcal{L} = ag_{a\gamma\gamma} F_{\mu\nu}\tilde{F}^{\mu\nu}/4$ $\,$         \\ \cline{2-3}
relics into $e^+e^-/\gamma\gamma$&	Dark photon, $A'$  	&	$\mathcal{L} = -\epsilon e A_\mu' J_{\rm em}^\mu$ $\,$         \\ \hline\hline
\end{tabular}
\caption{Summary of the scenarios that we identify as most promising to give a value of $N_{\rm eff}$ as low as $\sim 2.8$ and the specific models that realize them and for which we explicitly work out the phenomenology in Sections~\ref{sec:thermaleq} and~\ref{sec:nonthermalcase}.}
\label{tab:models}
\end{table*}

\section{Electromagnetic energy injections: The thermal case}\label{sec:thermaleq}

In this Section, we study thermal energy injections. Out-of-equilibrium injections are discussed in Section~\ref{sec:nonthermalcase}. We will consider explicit, simple, well-motivated models featuring this phenomenology.

\subsection{The Model}

Particles in thermal equilibrium during BBN have been studied in detail in the literature~\cite{Kolb:1986nf,Serpico:2004nm,Ho:2012ug,Boehm:2013jpa,Nollett:2013pwa,Sabti:2019mhn,Depta:2019lbe,Giovanetti:2021izc,Chu:2022xuh}. We revisit this scenario with a specific goal: to identify which particles can produce $N_{\rm eff}\simeq 2.8$ while remaining consistent with BBN in light of the new $Y_{\rm P}$ measurement in Eq.~\eqref{eq:Ypmeasurement}. A particularly appealing setup corresponds to a two-particle system made of a stable complex scalar $\phi$ and a dark photon $A'$, both with MeV-scale masses and with $m_{A'}>2m_\phi$, see e.g.~\cite{Feng:2017drg,Berlin:2018bsc}. $\phi$ will be a thermal dark matter relic, and its freeze-out abundance will be dictated by annihilations into $e^+e^-$ mediated by a virtual $A'$ that interacts with the Standard Model via kinetic mixing with the photon. 

This model is attractive because it avoids several important constraints. In particular, CMB bounds on annihilating dark matter are evaded as the annihilation is p-wave suppressed~\cite{Slatyer:2015jla}. If $\phi$ constitutes the dark matter, the scenario can be probed through $\phi e\to \phi e$ scattering in direct-detection experiments such as SENSEI~\cite{SENSEI:2019ibb,SENSEI:2019ibb}, DAMIC-M~\cite{DAMIC-M:2025luv,Castello-Mor:2020jhd}, and Oscura~\cite{Oscura:2023qik}; see~\cite{Krnjaic:2025noj,deBlas:2025gyz}. For $m_{A'}>2m_\phi$, the dark photon decays invisibly. This greatly weakens laboratory bounds, since invisible decays are harder to probe, and values as large as $\epsilon\sim 10^{-5}$ can remain compatible with existing searches; see e.g.~\cite{NA64:2023wbi,deBlas:2025gyz,NA64:2025ddk}. Such a large kinetic mixing also implies efficient production in supernova cores, where thermal equilibrium is established, and particles are trapped. This is expected not to modify the cooling time of core-collapse supernovae~\cite{Burrows:1990pk}, and at present no supernova bound excludes this region of parameter space~\cite{Chang:2018rso,Fiorillo:2025yzf}.

To obtain the right relic abundance, an interplay between the dark gauge-coupling $g_D$ and the kinetic mixing needs to take place, and this leads to further constraints. In the tens of MeV window for $\phi$ that we are interested in, dark matter self-interactions in the model may not be negligible. Requiring the self-scattering cross section per unit dark-matter mass to satisfy the Bullet Cluster bound, $\sigma_{\rm self}/m_\phi \lesssim {\rm cm^2}/{\rm g}$, implies $g_D \lesssim 0.9$~\cite{Tulin:2017ara,Randall:2008ppe,Sagunski:2020spe}. This, together with the requirement of the correct relic abundance and the bounds from NA64~\cite{NA64:2023wbi} on invisibly decaying dark photons, implies:
\begin{align}
    m_{A'}/m_\phi \lesssim 2.7\,\,\,\,[{\rm Bullet\, cluster+NA64}]\,.
\end{align}
Furthermore, a fairly similar limit can be cast from the lack of observed signals at direct detection dark matter experiments~\cite{Krnjaic:2025noj}. All in all, the available window for particles with masses in the tens of MeV is $2m_\phi\lesssim m_{A'} \lesssim 2.7m_\phi$, and we will focus our attention on it. With what respects to the cosmology, the $A'$ interacts extremely efficiently in the early Universe with $e^+e^-$ pairs and its partial decay width $\Gamma_{A'}$ satisfies $\Gamma_{A'}/H\sim 10^{10} \,(\epsilon/10^{-5})^2$, making the entire system of $\phi$ and $A'$ coupled to $e^+e^-$ and $\gamma$ particles. We find it is quite remarkable that such a simple model may have simultaneous implications for $N_{\rm eff}$, dark matter self-interactions, missing energy searches at accelerators, and dark matter direct detection experiments.

\begin{figure*}[!t]
\centering
\begin{tabular}{cc}
\hspace{-0.cm}\includegraphics[width=0.49\textwidth]{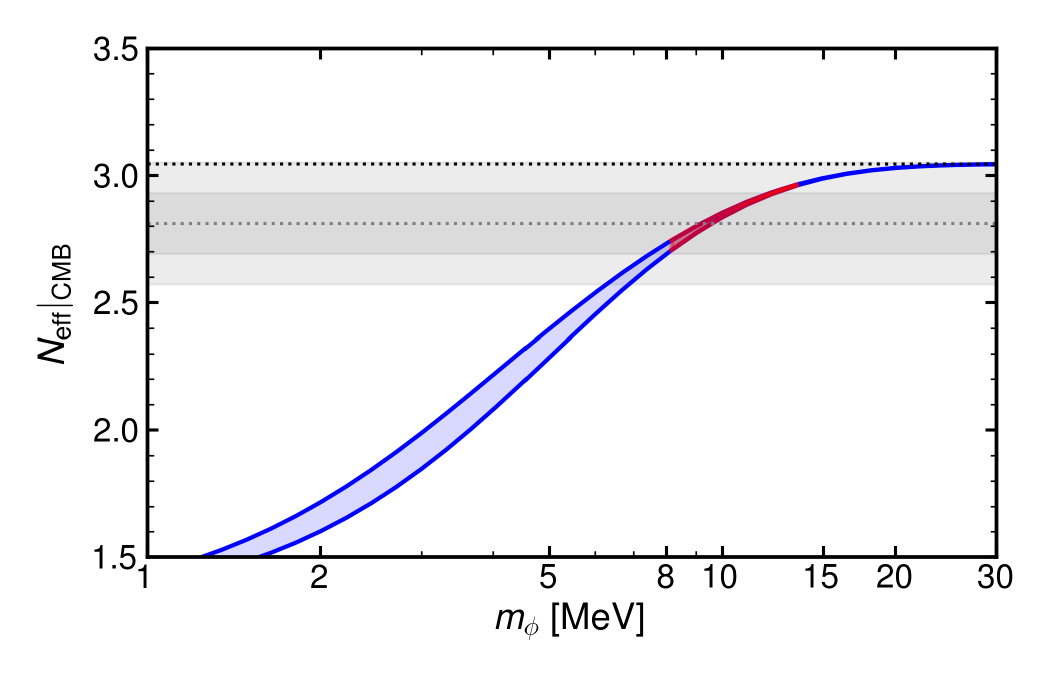}
&\hspace{-0.cm}\includegraphics[width=0.49\textwidth]{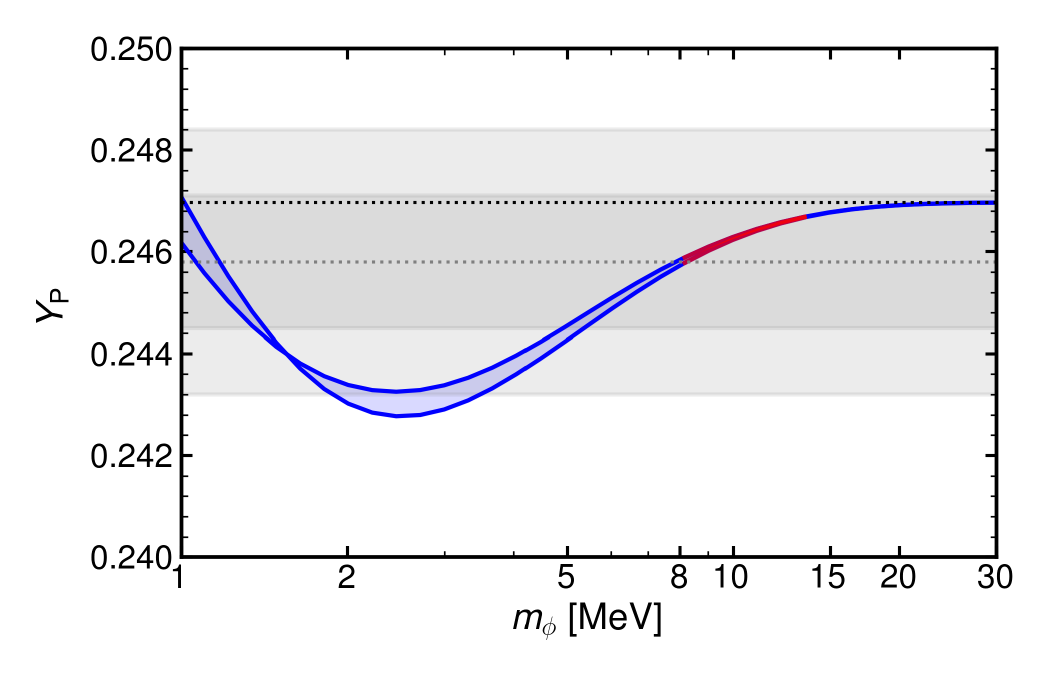} \\
\hspace{-0.cm}\includegraphics[width=0.49\textwidth]{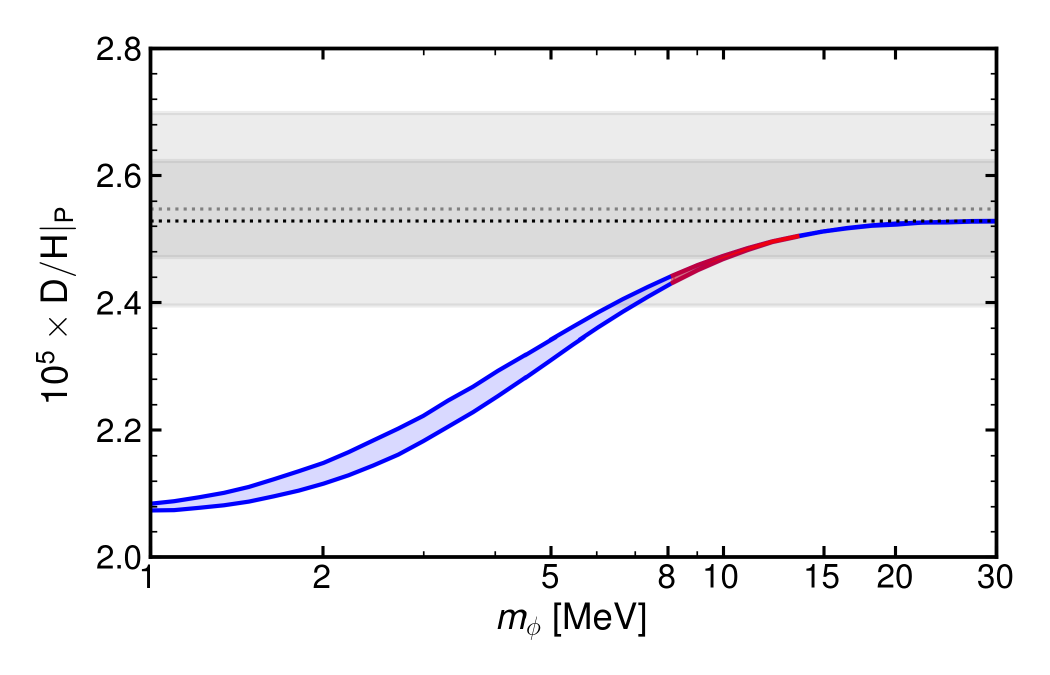} 
&\hspace{-0.cm}\includegraphics[width=0.49\textwidth]{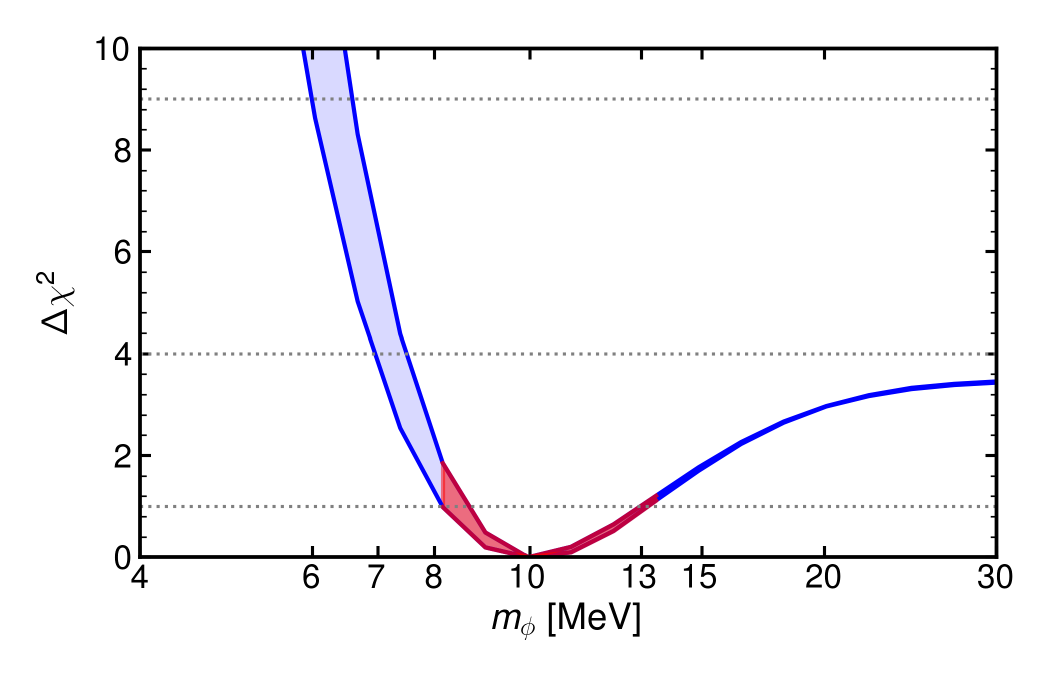} 
\end{tabular}
\vspace{-0.4cm}
\caption{
$N_{\rm eff}$, $Y_{\rm P}$ and ${\rm D/H}|_{\rm P}$ for the case of a thermal complex scalar $\phi$ interacting with a massive dark photon $A'$ with mass $m_{A'} = 2.7m_\phi$ (upper blue lines) and $m_{A'} = 2.01\,m_\phi$ (lower blue lines). The gray lines indicate the $1\sigma$ and $2\sigma$ confidence levels from observations, and for ${\rm D/H}|_{\rm P}$ we have added the theory uncertainty in quadrature to the measurement error. The lower-right panel shows the combined $\chi^2$ from these three observables (using $\Omega_bh^2$ from CMB observations). The minimum of the $\chi^2\simeq 1.4$ and in red we highlight the $\sim 1\sigma$ preferred region $m_\phi = 8-13\,{\rm MeV}$. This region is statistically preferred over $\Lambda$CDM only at the $\sim 1\sigma$ level. 
 }
\label{fig:thermal}
\end{figure*}

\subsection{The Cosmology}

With this in mind, we can easily solve for the early Universe thermodynamics. We will follow~\cite{Escudero:2018mvt,EscuderoAbenza:2020cmq,Escudero:2025kej} and consider all species to be described by thermal equilibrium distributions, and we will consider $T_\phi = T_{A'} = T_e = T_\gamma$ but $T_\nu \neq T_\gamma$. We assume that electromagnetic interactions are extremely efficient. It then suffices to solve for the evolution of the neutrino and photon temperatures, while accounting for energy injection and energy exchange between the electromagnetic ($\gamma$) and neutrino ($\nu$) sectors. The evolution equations for these temperatures are fairly simple and read:
\begin{subequations}\label{eqs:thermal}
\begin{align}
\frac{dT_\nu}{dt} = &-H \,T_\nu +\frac{Q_{\nu\, \leftarrow \,{\rm em}}}{4\rho_\nu/T_\nu}\,,\\
    \frac{{\rm d}T_{\gamma}}{{\rm d}t}  
= &- \bigg[4 H \rho_{\gamma} + 3 H \left( \rho_{e} + p_{e}\right)+ 3 H \left( \rho_\phi + p_\phi\right) \\
  &+ 3 H \left( \rho_{A'} + p_{A'}\right)  + 3 H  \, T_\gamma
  \frac{d p_{\rm int}}{dT_\gamma} + Q_{{\nu \,\leftarrow\, {\rm em}}}\bigg]  \nonumber  \\
  &\times \bigg[\frac{d \rho_{\gamma}}{d T_\gamma}
  + \frac{d \rho_e}{d T_\gamma}+ \frac{d \rho_\phi}{d T_\gamma}+ \frac{d \rho_{A'}}{d T_\gamma} + T_\gamma \frac{d^2 p_{\rm int}}{d T_\gamma^2}  \bigg]^{-1} \nonumber  \,,
\end{align}
\end{subequations}
where here $\rho$ and $p$ refer to energy densities and pressure of the various species, $H$ is the Hubble rate, $Q_{\nu \,\leftarrow\, {\rm em}}$ represents the energy transfer rate from the electromagnetic sector of the plasma to the neutrinos, and $p_{\rm int}$ represents the QED corrections to the pressure. To solve for these equations, we will use the publicly available \href{https://github.com/MiguelEA/nudec_BSM}{\texttt{\color{black} nudec\_BSM\_v2}} code 
and we employ the newly calculated rates and QED corrections from~\cite{Escudero:2025kej}.
 
To explore the BBN phenomenology, we implement these equations in our \textsc{Mathematica} BBN code. The code uses the main nuclear reaction rates as tabulated in the \textsc{PRIMAT} BBN code repository~\cite{Pitrou:2018cgg,Pitrou:2020etk}. We implement the proton-to-neutron conversion rates in the Born approximation and then add radiative corrections using a scaling function as derived in the Standard Model, which is accurate enough for our purposes. With this, we are able to simultaneously calculate $N_{\rm eff}$, $Y_{\rm P}$ and ${\rm D/H}|_{\rm P}$ as a function of $m_\phi$, $m_{A'}/m_\phi$ and $\Omega_b h^2$. For the latter, we will use $100\,\Omega_b h^2 = 2.238\pm 0.009$ as obtained within the $\Lambda$CDM model from Planck+ACT+SPT CMB observations~\cite{SPT-3G:2025bzu}. 
For ${\rm D/H}|_{\rm P}$ we will use the PDG recommended value of $10^5 \times {\rm D/H}|_{\rm P} =  2.55 \pm 0.03$~\cite{ParticleDataGroup:2024cfk}. In addition, we shall use the deuterium burning rates as in the \textsc{PArthENoPE} BBN code~\cite{Pisanti:2020efz,Gariazzo:2021iiu}. This leads to a ${\rm D/H}|_{\rm P}$ prediction compatible with the observed one, and implies a theoretical error in the prediction of $\pm 0.07 \times 10^{-5}$ for ${\rm D/H}|_{\rm P}$\footnote{We should note that the deuterium burning rates output in~\cite{Pitrou:2018cgg,Pitrou:2020etk} are also perfectly valid ones, but that they lead to a deuterium abundance that falls $\sim 2\sigma$ low compared to the measured value. We elaborate upon the potential implications of this result in the conclusions.}. 

The results from our analysis are shown in Fig.~\ref{fig:thermal}. It highlights that only a narrow mass window is favored over $\Lambda$CDM. It corresponds to $m_\phi=8$--$13\,{\rm MeV}$ and $m_{A'}=(2$--$2.7)\,m_\phi$. This represents a clear target for dark matter direct detection via dark matter-electron scatterings. In addition, for moderately large couplings, the dark photon in the model can also be searched for with beam-dump experiments such as NA64, see Fig. 9.4 of~\cite{deBlas:2025gyz}. We note that our analysis shows that $m_\phi < 6\,{\rm MeV}$ is strongly disfavored by BBN and CMB observations. 

\begin{figure*}[!t]
\centering
\begin{tabular}{cc}
\hspace{-0.cm}\includegraphics[width=0.49\textwidth]{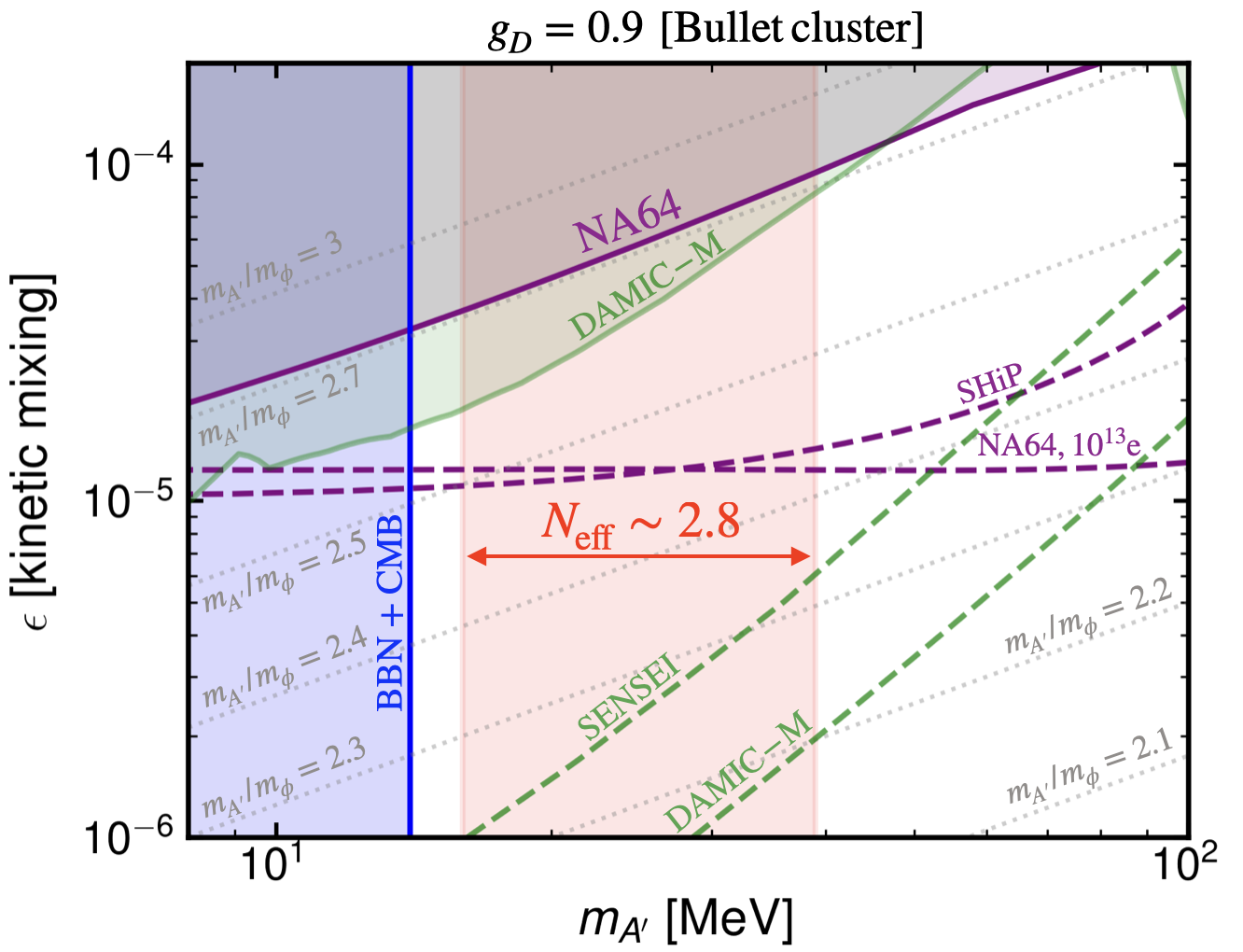}
&\hspace{-0.cm}\includegraphics[width=0.49\textwidth]{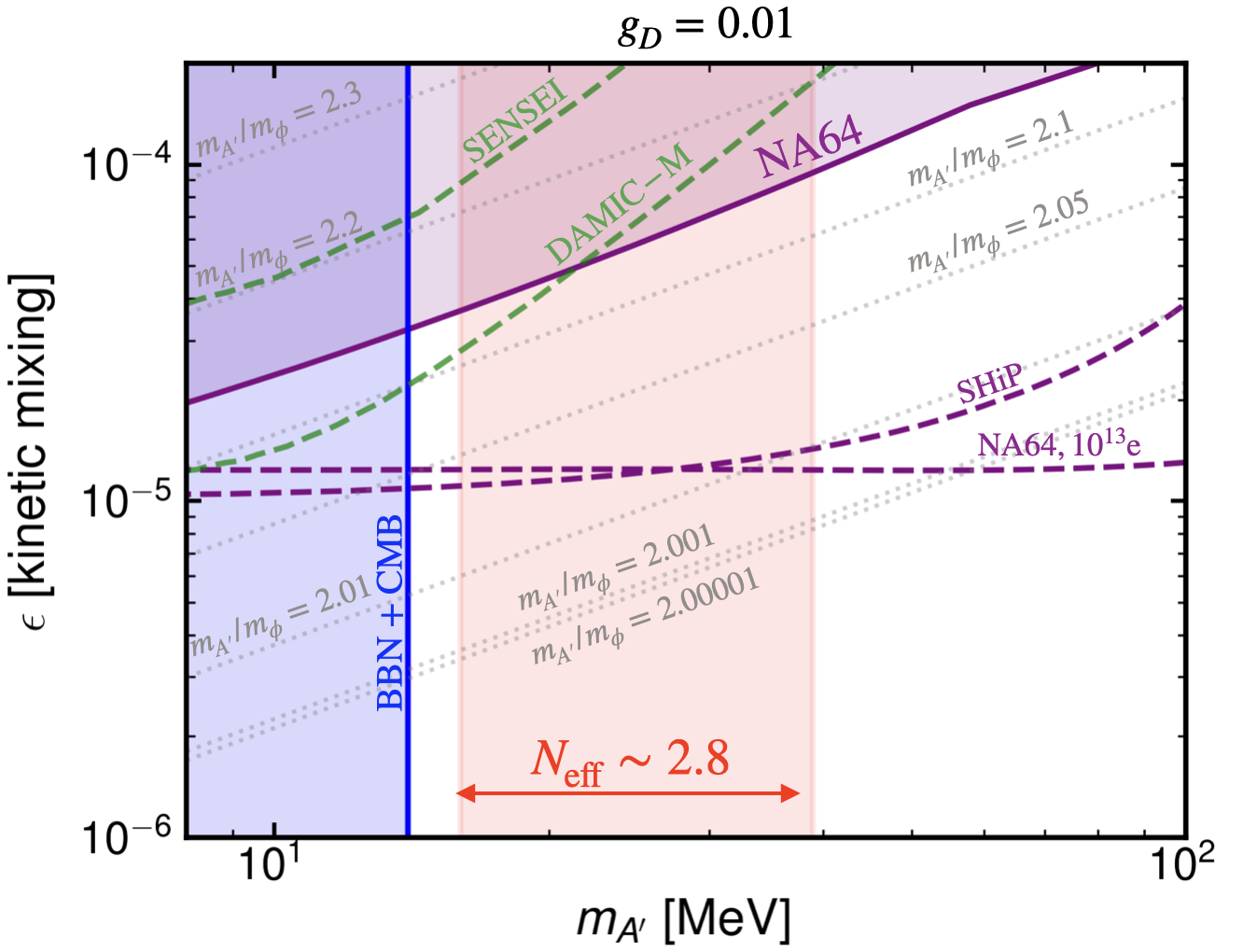}
\end{tabular}
\vspace{-0.4cm}
\caption{
Parameter space of the thermal scenario iia) of Table~\ref{tab:models} for two values of $g_D$: $g_D = 0.9$, the largest value allowed by the Bullet Cluster bound (left), and $g_D=0.01$ (right). The gray dotted lines show, for the indicated value of $m_{A'}/m_\phi$, the kinetic mixing at which $\phi$ reproduces the observed dark matter abundance. The red band marks the dark photon masses for which $N_{\rm eff}\simeq 2.8$, and the blue region marks the masses excluded by the BBN and CMB analysis of Fig.~\ref{fig:thermal}. The remaining colored regions are excluded by the laboratory searches indicated, and the dashed lines are the lower boundaries of the regions that laboratory searches are projected to reach within the next two decades~\cite{deBlas:2025gyz}. The NA64 constraints and sensitivities are taken from Ref.~\cite{NA64:2025ddk}, DAMIC-M from Ref.~\cite{DAMIC-M:2025luv}, and SHiP from Refs.~\cite{Kyselov:2024dmi,deBlas:2025gyz}. The direct detection limits and projections are quoted as a function of $m_\phi$, and to draw them here we use $m_{A'} = 2.5\,m_\phi$; choosing instead $m_{A'} = (2-2.7)\,m_\phi$ does not change them significantly.
 }
\label{fig:bounds}
\end{figure*}

How does this model evade the strong bounds from $Y_{\rm P}$ while leading to a fairly sizable impact on $N_{\rm eff}$? The reason is a partial cancellation between two effects affecting the helium synthesis. The helium abundance is dominated by the neutron one, as effectively all neutrons available at BBN bind into helium-4. In this model, the energy release by $\phi\bar{\phi}$ annihilations leads to a larger $T_\gamma/T_\nu$ than in the Standard Model at $T\sim 1\,{\rm MeV}$, which is when proton-to-neutron interactions freeze out. Here is where the two effects almost cancel. First, for fixed $T_\gamma$, the Hubble rate is lower than in the Standard Model, which tends to decrease the helium abundance. Second, the weak interaction rates are also smaller because they depend on the neutrino temperature, and this tends to increase it. As a result, $\Gamma_{p\leftarrow n}/H$ remains approximately unchanged, and the primordial helium abundance stays close to its Standard Model value. Importantly, since this happens at $T\sim 1\,{\rm MeV}$, the deuterium abundance, which builds up at $T\sim 0.075\,{\rm MeV}$, is strongly correlated to the value of $N_{\rm eff}$ as seen in Fig.~\ref{fig:thermal}.

This scenario can be tested simultaneously in missing energy searches for invisibly decaying dark photons, in dark matter direct detection experiments, and through its impact on dark matter self-interactions. We show this synergy between cosmological and laboratory signatures in Fig.~\ref{fig:bounds}, for two values of $g_D$: the largest value allowed by the Bullet Cluster bound, $g_D\simeq 0.9$, where direct detection experiments provide the strongest sensitivity, and a smaller coupling, $g_D = 0.01$, where accelerator searches dominate the sensitivity.

\vspace{-0.6cm}

\section{Electromagnetic out-of-equilibrium decays} \label{sec:nonthermalcase}

In this section, we consider a particle $X$ that decays out of equilibrium into electromagnetic radiation, $X\to e^+e^-$ or $\gamma\gamma$, at times near neutrino decoupling, $t_U \sim 1\,{\rm s}$ and $T\sim {\rm MeV}$. As summarized in Table~\ref{tab:models}, there are two simple states beyond the Standard Model that feature this phenomenology: one are axion-like particles decaying to photons $a\to \gamma \gamma$, and the other are dark photons decaying into electron-positron pairs $ A'\to e^+e^-$.

\begin{figure*}[!t]
\centering
\begin{tabular}{cc}
\hspace{-0.cm}\includegraphics[width=0.49\textwidth]{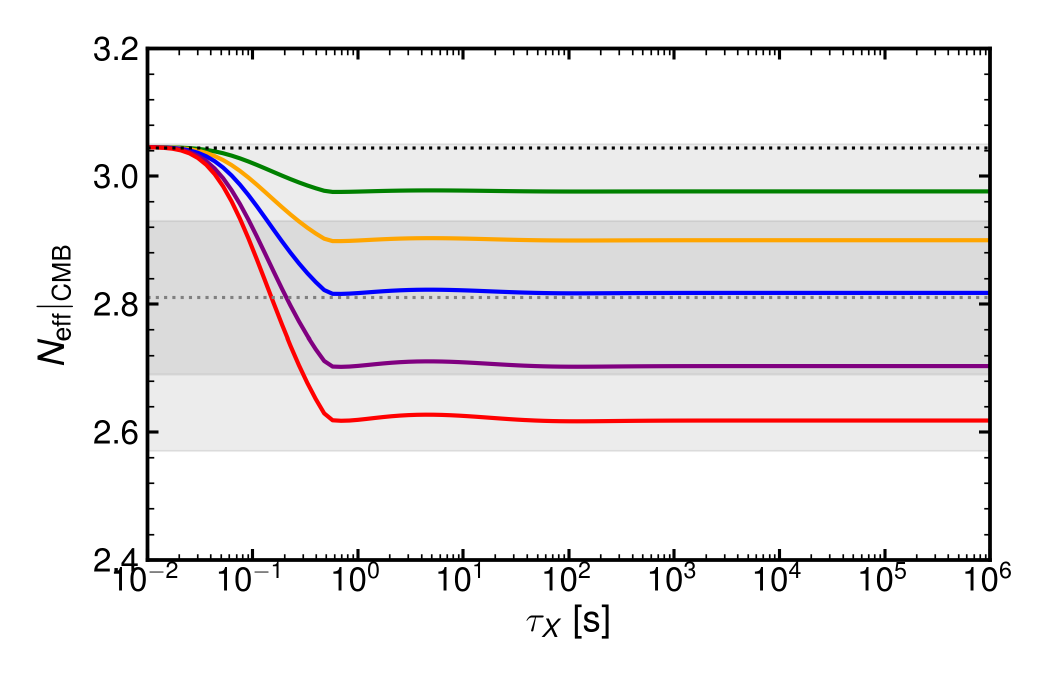}
&\hspace{-0.cm}\includegraphics[width=0.49\textwidth]{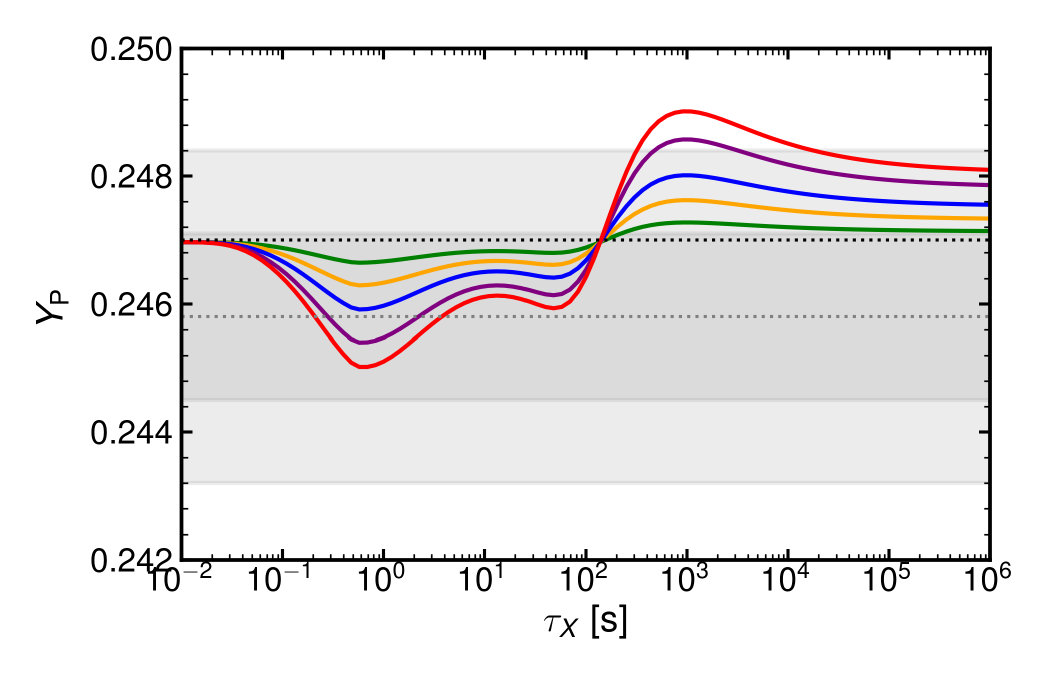} \\
\hspace{-0.cm}\includegraphics[width=0.49\textwidth]{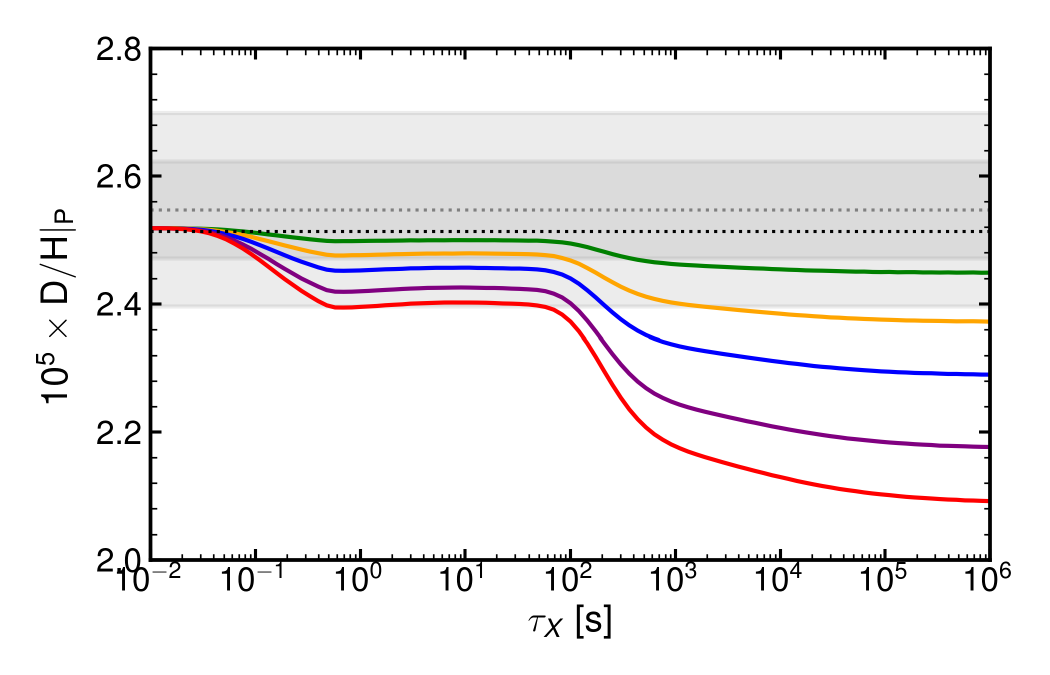} 
&\hspace{-0.cm}\includegraphics[width=0.49\textwidth]{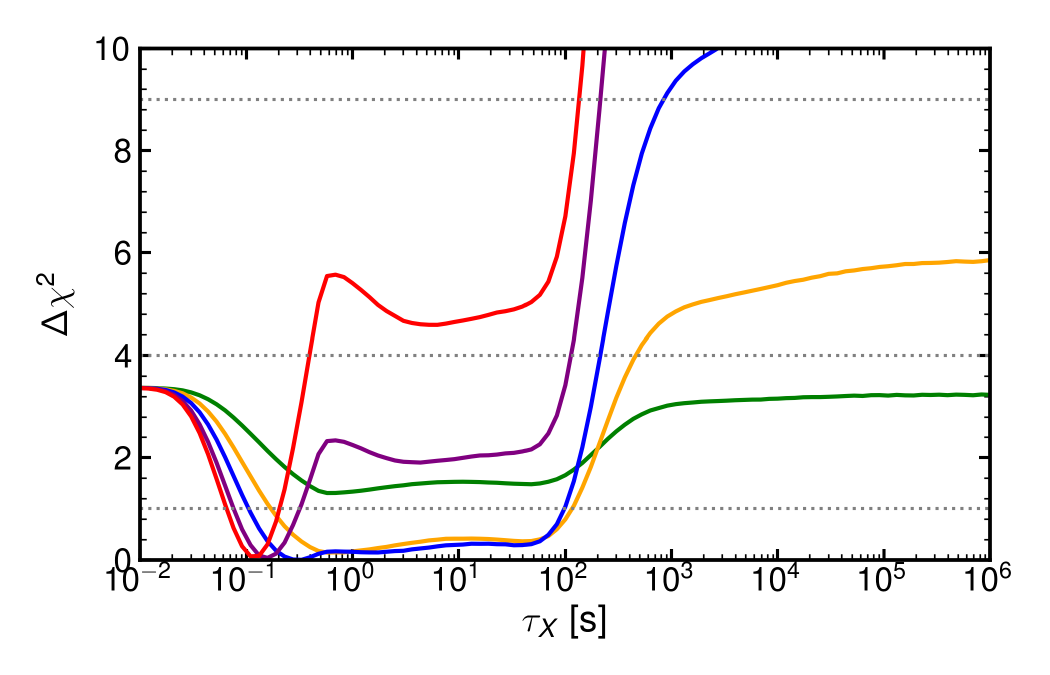} 
\end{tabular}
\vspace{-0.6cm}
\caption{
$N_{\rm eff}$, $Y_{\rm P}$ and ${\rm D/H}|_{\rm P}$ for the case of a relic decaying out-of-equilibrium to the EM particles $X\to e^+e^-$ or $X\to \gamma\gamma$, as a function of the relic's lifetime $\tau_X$. The particular value of the relic mass is irrelevant as long as $m_{X}<2m_{\mu}$, as these electromagnetic decay products thermalize in the plasma. We show colored lines for certain values of the number density abundance before decay, fixed in a way such that $N_{\rm eff}$ tends to a constant value in the limit $\tau_X \gtrsim 1\text{ s}$. In the region $0.05\,{\rm s}\lesssim \tau_X \lesssim 100\,{\rm s}$ these types of particles are statistically favored over $\Lambda$CDM, although again only at the $\sim 1\sigma$ level. The lower right panel refers to $\Delta\chi^2 = \chi^2-\chi^2|_{\rm min}|_{\rm blue}$ where $\chi^2|_{\rm min}|_{\rm blue}$ is used as it is the scenario with the lowest value of $\chi^2$.
 }
\label{fig:outofeq}
\end{figure*}

For the moment, we remain agnostic about the underlying particle model. To describe the early-Universe thermodynamics, we again follow~\cite{Escudero:2018mvt,EscuderoAbenza:2020cmq,Escudero:2025kej} and assume thermal distributions for $e^+e^-$, photons, and neutrinos, with $T_e=T_\gamma$ but $T_\nu\neq T_\gamma$. Since the electromagnetic energy injections will thermalize very efficiently, this represents an accurate description of the early Universe thermodynamics, and the phenomenology of $e^+e^-$ and $\gamma \gamma$ injections is identical. The evolution equations that dictate the evolution of the system are:
\begin{subequations}\label{eq:nonthermal}
\begin{align}
\frac{dT_\nu}{dt} = &-H \,T_\nu +\frac{Q_{\nu\, \leftarrow \,{\rm em}}}{4\rho_\nu/T_\nu}\,,\\
    \frac{{\rm d}T_{\gamma}}{{\rm d}t}  
= &- \bigg[4 H \rho_{\gamma} + 3 H \left( \rho_{e} + p_{e}\right)- \Gamma_X \rho_X \\
  & +3 H  \, T_\gamma
  \frac{d p_{\rm int}}{dT_\gamma} + Q_{{\nu \,\leftarrow\, {\rm em}}}\bigg] \nonumber \\
  &\times \bigg[\frac{d \rho_{\gamma}}{d T_\gamma}
  + \frac{d \rho_e}{d T_\gamma} + T_\gamma \frac{d^2 p_{\rm int}}{d T_\gamma^2}  \bigg]^{-1}  \,,\nonumber \\
 \frac{d \rho_X}{dt} &=-  3H \rho_X  -\Gamma_X \rho_X\,,
\end{align}
\end{subequations}
where the notation is as in Eq.~\eqref{eqs:thermal} and where $\Gamma_X = 1/\tau_X$ is the decay rate of the $X$ particle. We have implicitly assumed that the mass of this state is $m\gg 10\,{\rm MeV}$ and considered ${\rm BR} = 1$ into $e^+e^-/\gamma\gamma$. 

If the energy density of $X$ is never a substantial fraction of the energy density in the Standard Model, then its evolution equation can be integrated explicitly, see e.g.~\cite{Kolb:1990vq}. For a radiation dominated Universe $H = 1/(2t)$ and then it is easy to see that:
\begin{align}
   \!\!\!\! \rho_X(t) = \rho_X^0(t_0) a^{-3} e^{-t/\tau_X} = \rho_X^0(t_0) (t/t_0)^{-3/2} e^{-t/\tau_X}\,,
\end{align}
where $t_0$ is some arbitrary initial time and $a$ is the scale factor. The beauty of this equation is that it allows us to integrate it to estimate the contribution to $\Delta N_{\rm eff}$. We will have $\delta N_{\rm eff} \simeq -3\,\delta\rho_\gamma/\rho_\gamma  $ where $\delta\rho_\gamma$ is approximately the energy density injected into electromagnetic radiation from $X$ decays. This should be seen as an integral and simply reads:
\begin{align}
    \Delta N_{\rm eff} \propto -3\int_{0}^{\infty} dt \, \Gamma_X\, \frac{\rho_X}{\rho_\gamma+\rho_e}\,.
\end{align}
We can use that $\rho_\gamma+\rho_e = 2 \pi^2/30 T_\gamma^4 (1+f_e)$ where $f_e = 7/4$ at $T\gg m_e$ and $f_e = 0$ at $T\ll m_e$. We can also approximately trade $T_\gamma$ with time using the Hubble equation to find: $T_\gamma^2 \simeq M_{\rm Pl}/(2t\times 1.66\sqrt{g_\star}) $ where $g_\star \simeq 10.75$. With this, we can do the integral explicitly and find an approximate relationship:
\begin{align}
    \Delta N_{\rm eff} \propto - \frac{\rho_X(\tau_X)}{\rho_\gamma(\tau_X)}\,.
\end{align}
In other words, $\Delta N_{\rm eff}$ is proportional to the ratio of the $X$ energy density to the photon energy density at the time of decay, where the constant of proportionality is $\simeq 1/6-1/5$. To solve for the thermodynamics, we start the system at $T_\gamma =T_\nu = 10\,{\rm MeV}$ and with a given initial condition for $\rho_X$. 

Our results are shown in Fig.~\ref{fig:outofeq}. Each colored line corresponds to an initial condition for $\rho_X$ leading to approximately the same energy density of the $X$ particle at the time of its decay. As anticipated in Section~\ref{sec:globalconsiderations}, the preferred injection window is $0.05\,{\rm s}\lesssim \tau\lesssim 100\,{\rm s}$. At times earlier than these, the electromagnetic radiation thermalizes with neutrinos, and $\neff \simeq N_{\rm eff}|_{\rm SM}$. For $t\gtrsim 100\,\s\sim t_{\rm BBN}$, the decay dilutes the baryon density and produces strong tension with the primordial deuterium abundance; see the lower-left panel. Since the injection of energy occurs roughly after proton-to-neutron freeze-out at $T\lesssim 0.7\,{\rm MeV}$, the helium abundance is not impacted significantly.

From the lower right panel of Fig.~\ref{fig:outofeq}, we can clearly see that the blue contour gives some of the best-fitting regions across parameter space. This region roughly corresponds to a particle with a primordial abundance of:
\begin{align}
\label{eq:Xprimordialabundance}
    Y_X \equiv n_X/s \simeq 8\times 10^{-4} \left(\frac{250\text{ MeV}}{m_X}\right) \left(\frac{\rm 10\,{\rm s}}{\tau_X}\right)^\frac{1}{2}\,,
\end{align}
where $s$ is the entropy density of the Universe. We note that if the lifetime of the particle is $\tau_X\sim 0.1\,{\rm s}$, its abundance may be larger.

\textbf{On the mass of the out-of-equilibrium decaying particle:} Not only does the early Universe phenomenology require the lifetime of the $X$ particle to lie in a rather specific window of $0.05\,{\rm s} \lesssim \tau_X\lesssim 100\,{\rm s}$, but the following two considerations also bound quite strongly its mass. A particle with such lifetimes is copiously produced in supernova cores~\cite{Raffelt:1996wa,Caputo:2024oqc}. This is the case, for instance, for an axion-like particle decaying into a pair of photons $a\to \gamma \gamma$ where an array of astrophysical bounds arise~\cite{Chang:2018rso,Caputo:2022mah,Diamond:2023cto,Diamond:2023cto,Jaeckel:2017tud,Caputo:2021rux,Hoof:2022xbe,Candon:2025ypl,Fiorillo:2025yzf} and similarly for a dark photon~\cite{Redondo:2008ec,Chang:2016ntp,Caputo:2025avc,Caputo:2026pdw}. These bounds, however, are only effective if the particle can be produced in the core of supernovae or neutron star mergers. Given the temperature of the supernova $T\sim 30\,{\rm MeV}$, particles with large masses are simply not produced. Indeed, if the mass of this particle is above $\sim 250-300\,{\rm MeV}$, these limits are evaded.

However, another issue appears precisely for these large masses: the particle will start decaying into mesons as it is kinematically unavoidable. For the axion-like particles, this will happen as a result of off-shell photon-mediated decays, e.g., $a\to \gamma \gamma^*\to \gamma \pi^+\pi^-$, while for the dark photon, this happens directly from its mixing with the photon, and the branching ratio is typically much larger. The effect of mesons is potentially catastrophic, as even tiny amounts of mesons may strongly interact with nuclei, leading to an excess synthesis of both helium and deuterium in the Universe, see~\cite{Reno:1987qw,Kawasaki:2000en,Pospelov:2010cw,Akita:2024ork}. This issue was recently investigated by some of us in Ref.~\cite{Escudero:2025avx} in the context of an axion-like particle that couples exclusively to photons. If its mass is above $m\gtrsim 600\,{\rm MeV}$, the BBN bounds resulting from the effect of mesons will not allow a significant modification of $N_{\rm eff}$. Quite remarkably, not only the lifetime of this putative particle would be constrained but also its mass: $250\,{\rm MeV}\lesssim m_a\lesssim 600\,{\rm MeV}$. For the dark photon case, the branching ratio into hadrons is much larger, and the combination of these two issues would require $250\,{\rm MeV}\lesssim m \lesssim 300\,{\rm MeV}$. 

\textbf{On the abundance of the $X$ particle:} The required primordial abundance of the $X$ particle in Eq.~\eqref{eq:Xprimordialabundance} is another ingredient in the model, and it would be very appealing if the same interactions that trigger the $X$ decay also lead to the right primordial population in the early Universe. This is not the case for dark photons~\cite{Redondo:2008ec} as their primordial production is primarily dominated by inverse decays $e^+e^-\to A'$, which produce negligible abundance for a particle with $m_X \sim 250\,{\rm MeV}$ and $\tau_X\sim 10\,{\rm s}$. The abundance of an axion-like particle is dominated by Primakoff interactions $e \gamma \to ea$. For an ALP in a purely radiation-dominated Universe, it may actually lead to a primordial thermal abundance of these states which is much larger than Eq.~\eqref{eq:Xprimordialabundance}. On the other hand, if there is some entropy dilution or the reheating temperature of the Universe is $T\sim 10^{6}\,{\rm GeV}$, one could get the right ALP abundance~\cite{Escudero:2025avx}. It turns out that having the right abundance requires extra ingredients in both of these models.

\section{Discussion \& Conclusions}
\label{sec:conclusions}

We are now firmly in the era of precision cosmology. Within the past year, the radiation content of the Universe has been constrained with unprecedented precision, both directly from CMB data and indirectly from BBN through a new high-precision measurement of the primordial helium abundance. The fact that these measurements agree to the degree they do, even while arising from dramatically different eras of cosmic history, is remarkable. 

As we look to the future, the recently deployed Simons Observatory is expected to deliver unprecedented precision on \neff. However, the pull of existing CMB data with a central value $N_{\rm eff}\sim 2.8$ makes the prospect of a strong detection of excess radiation ($\Delta N_{\rm eff}>0$) unlikely, at least in the most minimal extensions of $\Lambda$CDM. In this context, it is timely and interesting to understand what extensions of the Standard Model can yield $N_{\rm eff}<3.044$. In this study, we have precisely explored such models and their compatibility with all other cosmological, astrophysical, and laboratory data, in particular given the very recent measurement of the primordial helium abundance.

In Section~\ref{sec:globalconsiderations}, we considered a wide variety of BSM options featuring $N_{\rm eff}<3$ and discussed the cosmological constraints they face. We focused in particular on scenarios that can lead to $N_{\rm eff}\simeq2.8$, as summarized in Fig.~\ref{fig:summary}. We find that two particularly simple scenarios can achieve this:
\begin{itemize}
    \item[iia)] Thermal scalar dark matter $\phi$ with mass:
    \begin{align}
        m_{\phi} \in (8-13)\,{\rm MeV}\,,
    \end{align}
    accompanied by a kinetically mixed dark photon $A'$ with mass
    \begin{align}
        2\, m_\phi\lesssim m_{A'}\lesssim 2.7\, m_\phi\,.
    \end{align} 
    As discussed in Section~\ref{sec:thermaleq}, due to a compensation of effects, this scenario features a helium abundance that is Standard Model-like. Quite remarkably, this rather simple model can be simultaneously tested via: 1) searches for invisibly decaying dark photons at experiments such as NA64~\cite{NA64:2025ddk} or LDMX~\cite{LDMX:2025bog}, 2) $\phi e\to \phi e$ scatterings at direct detection experiments such as SENSEI, DAMIC-M, and Oscura, see~\cite{Krnjaic:2025noj}, 3) its impact on dark matter structures from non-negligible self-interactions, and 4) by the Simons Observatory from $N_{\rm eff}$ measurements. The interplay of these probes is shown in Fig.~\ref{fig:bounds}.
    \item[iib)] A particle $X$ decaying out-of-equilibrium into $e^+e^-/\gamma\gamma$ with the following properties:
    \begin{align}
 \,\,\,     &0.05 \lesssim \,\tau_X/\,{\rm s}\lesssim 100\,    \,\,\,[{\rm BBN}]\,,\\
[{\rm SN}]  \,\,\, \,\,    &250 \lesssim \,\frac{m_X}{\rm MeV}\lesssim 600   \,\,\,[{\rm BBN}]\,,\nonumber \\
Y_X \simeq  8\times 10^{-4}& \left(\frac{250\text{ MeV}}{m_X}\right)\! \left(\frac{\rm 10\,{\rm s}}{\tau_X}\right)^\frac{1}{2}\,\,\,[N_{\rm eff}\simeq 2.8]\nonumber\,,
    \end{align}
    where we have indicated the origin of the bound/requirement as elaborated in Section~\ref{sec:nonthermalcase}. This scenario evades the strong $Y_{\rm P}$ bounds because the injection happens after proton-neutron conversions freeze out, and two types of simple BSM models feature this phenomenology: photophilic axion-like particles and dark photons, see Table~\ref{tab:models}. 
\end{itemize}
Remarkably, these very simple, specific scenarios can yield a significantly lower $N_{\rm eff}$ than 3.044 at CMB times and remain compatible with all known laboratory, astrophysical, and cosmological data. 

\textit{Deuterium as a key observable:} As can be seen from the lower-left panels of both Fig.~\ref{fig:thermal} and Fig.~\ref{fig:outofeq}, the deuterium abundance is a key observable to test these scenarios where electromagnetic radiation is injected before Big Bang Nucleosynthesis, as its abundance is strongly correlated with $N_{\rm eff}$. In these setups, the expansion rate of the Universe is lower than in the Standard Model, which leads to a smaller deuterium abundance (because more deuterium can burn). At present, the situation with the deuterium abundance ${\rm D/H}$ is a bit delicate. It has been measured with very high precision and with small systematics from quasar absorption~\cite{Cooke:2017cwo}. However, its theoretical prediction is quite sensitive to yet not very well known reaction rates: ${\rm D+D\to T}+p$ and ${\rm D+D\to {}^3{\rm He}} + n$. Depending upon the modeling of these rates, various groups report different predictions for this abundance:
\begin{subequations}
\begin{align}
\!\! \!\!\! \!10^5  \,  {\rm D/H}|_{\rm P}^{\rm SM} &= 2.51\pm 0.07\,,~\cite{Pisanti:2020efz,Gariazzo:2021iiu}\,\,{\rm (Prediction)}\\
\!\! \!\!\! \!10^5  \,  {\rm D/H}|_{\rm P}^{\rm SM} &= 2.48\pm 0.08\,,~\cite{Yeh:2020mgl,Yeh:2022heq}\,\,{\rm (Prediction)}\\
\!\! \!\!\! \!10^5  \,  {\rm D/H}|_{\rm P}^{\rm SM} &= 2.44\pm 0.04\,, ~\cite{Pitrou:2018cgg,Pitrou:2020etk}\,\,{\rm (Prediction)}\label{eq:DHPRIMAT}\\
\!\! \!\!\! \! 10^5  \,  {\rm D/H}|_{\rm P}^{\rm Obs} &=  2.55 \pm 0.03\,.~\cite{ParticleDataGroup:2024cfk}\,\,{\rm (Measurement)}
\end{align}
\end{subequations}
It becomes clear that an accurate measurement of these reaction rates~\cite{Pitrou:2021vqr} can have an important impact and allow for a test of these scenarios. Furthermore, we note that the predictions of Refs.~\cite{Pitrou:2018cgg,Pitrou:2020etk} lie somewhat below the observed deuterium abundance. The scenarios considered here, which lead to $N_{\rm eff}<3$, generally reduce the predicted deuterium abundance even further and could therefore exacerbate any emerging tension. We conclude by remarking that if the electromagnetic injection comes with a small number ${\rm BR}\sim 10^{-5}$ of light mesons, they can, in principle, increase the deuterium abundance in a chain $X\to \pi^+\pi^-$, $p+\pi^- \to n + \pi^0$, and $n+p\to {\rm D}+\gamma$. This can happen, for example, for the axion-like particle if it has a mass $\sim 500\,{\rm MeV}$ and a lifetime $\sim 10^3\,{\rm s}$, see~\cite{Escudero:2025avx}. If the theoretical prediction for the deuterium abundance improves in the future and it is confirmed that it is significantly below the observed one, a particle with these properties may simultaneously address both effects. 

At present, we do not know whether $N_{\rm eff}$ is smaller than, equal to, or larger than three. In this study, we have focused on scenarios leading to $N_{\rm eff}<3$ and explicitly explored the phenomenology of models that are simple and perfectly viable. Upcoming high-precision cosmological data will determine whether these scenarios are favored or excluded.

\vspace{-0.3cm}
\section*{Acknowledgments}
\vspace{-0.3cm}
We are grateful to Graham Kribs and Ryan Plestid for pointing out the relevance of dark matter self-interactions in our scenario iia) and Maria Ramos for useful discussions on this phenomenology. We would also like to thank Andrea Caputo for useful discussions about supernova bounds. MO received support from the European Union's Horizon Europe research and innovation programme under the Marie Sklodowska-Curie grant agreement No~101204216. N.W. is supported by the National Science Foundation grant PHY-2210498, by the BSF under grant 2022287, and the Simons Foundation.

\bibliography{biblio}

\end{document}